\documentclass[journal,comsoc]{IEEEtran}
\usepackage{graphicx}
\usepackage{url}
\usepackage{cite}
\usepackage{color}
\usepackage{amsmath}
\usepackage{amssymb}
\usepackage{latexsym}
\usepackage[cmintegrals]{newtxmath}
\usepackage{algorithm}
\usepackage{algorithmic}
\usepackage{subfigure}
\usepackage{stfloats} % to put the equation at the bottom of page.
\usepackage{multirow} % use multiple rows in the table.

\usepackage[acronym]{glossaries}
\usepackage{xurl}
\newacronym{mMTC}{mMTC}{massive machine-type communications}
\newacronym{eMBB}{eMBB}{enhanced mobile broadband}
\newacronym{URLLC}{URLLC}{ultra reliable low latency communications}
\newacronym{IoT}{IoT}{Internet-of-Things}
\newacronym{CS}{CS}{compressed sensing}
\newacronym{DCS}{DCS}{distributed CS}
\newacronym{MIMO}{MIMO}{multiple-input-multiple-output}
\newacronym{NOMA}{NOMA}{non-orthogonal multiple access}
\newacronym{GFNOMA}{GF-NOMA}{grant-free non-orthogonal multiple access}
\newacronym{BCD}{BCD}{block coordinate descent}
\newacronym{CD}{CD}{coordinate descent}
\newacronym{PG}{PG}{proximal gradient}
\newacronym{FISTA}{FISTA}{fast iterative shrinkage-thresholding algorithm}
\newacronym{AUD}{AUD}{active user detection}
\newacronym{CE}{CE}{channel estimation}
\newacronym{MUD}{MUD}{multiuser detection}
\newacronym{JACE}{JACE}{joint activity and channel estimation}
\newacronym{JACDE}{JACDE}{joint activity, channel, and data estimation}
\newacronym{AUE}{AUE}{active user enumeration}
\newacronym{BS}{BS}{base station}
\newacronym{AMP}{AMP}{approximate message passing}
\newacronym{MMVAMP}{MMV-AMP}{multiple measurement vector approximate message passing}
\newacronym{CBMMVAMP}{CB-MMV-AMP}{completely-blind MMV-AMP}
\newacronym{GMMVAMP}{GMMV-AMP}{generalized MMV-AMP}
\newacronym{GAMP}{GAMP}{generalized approximate message passing}
\newacronym{iid}{i.i.d.}{independently and identically distributed}
\newacronym{AWGN}{AWGN}{additive white Gaussian noise}
\newacronym{5GNR}{5G NR}{fifth-generation new radio}
\newacronym{SNR}{SNR}{signal-to-noise ratio}
\newacronym{LSF}{LSF}{large-scale fading}
\newacronym{MD}{MD}{miss detection}
\newacronym{FA}{FA}{false alarm}
\newacronym{ML}{ML}{maximum likelihood}
\newacronym{MLE}{MLE}{maximum likelihood estimation}
\newacronym{EM}{EM}{expectation maximization}
\newacronym{NMSE}{NMSE}{normalized mean-squared error}
\newacronym{NRMSE}{NRMSE}{normalized root mean-squared error}
\newacronym{MMSE}{MMSE}{minimum mean-squared error}
\newacronym{MSE}{MSE}{mean-squared error}
\newacronym{AER}{AER}{activity error rate}
\newacronym{TO}{TO}{timing offset}
\newacronym{CFO}{CFO}{carrier frequency offset}

\begin{document}
\title{Eigenvalue Based Active User Enumeration for Grant-Free Access Under Carrier Frequency Offsets}

\author{Takanori~Hara~\IEEEmembership{Member,~IEEE}% <-this % stops a space
\thanks{This work has been submitted to the IEEE for possible publication. Copyright may be transferred without notice, after which this version may no longer be accessible. This work was partially supported by JSPS KAKENHI Grant Number JP23K13337 and Telecommunications Advancement Foundation Research Grant. T. Hara is with the Department of Electrical Engineering, Tokyo University of Science, Chiba 278-8510, Japan (email: t.hara@ieee.org).}
% \thanks{This work was partially supported by JSPS KAKENHI Grant Number JP23K13337 and Telecommunications Advancement Foundation Research Grant.}
% \thanks{T. Hara is with the Department of Electrical Engineering, Tokyo University of Science, Chiba 278-8510, Japan (email: t.hara@ieee.org).}
%\thanks{T. Hara is with the Department of Electrical Engineering, Tokyo University of Science, 2641 Yamazaki, Noda-shi, Chiba 278-8510, Japan (email: t.hara@ieee.org).}
%\thanks{Manuscript received xxxxx xx, xxxx; revised xxxxx xx, xxxx.}
}

\markboth{PREPRINT}%
{Shell \MakeLowercase{\textit{et al.}}: Bare Demo of IEEEtran.cls for IEEE Journals}
% *** name in the headers of peer review papers.  ***



\maketitle

\begin{abstract}
This paper investigates a grant-free non-orthogonal multiple access (GF-NOMA) system in the presence of carrier frequency offsets. 
We propose two schemes for enumerating active users in such a GF-NOMA system, which is equivalent to estimating the sparsity level. 
Both schemes utilize a short common pilot and the eigenvalues of the sample covariance matrix of the received signal. 
The two schemes differ in their treatment of noise variance: one exploits known variance information, while the other is designed to function without this knowledge. 
Simulation results demonstrate the effectiveness of the proposed schemes in terms of the \textcolor{black}{normalized root-mean-squared error}. 
\end{abstract}

% Note that keywords are not normally used for peer review papers.
\begin{IEEEkeywords}
Grant-free access, active user enumeration, sparsity level estimation, carrier frequency offsets. 
\end{IEEEkeywords}

% \fi
% creates the second title. It wi1l be ignored for other modes.
\IEEEpeerreviewmaketitle

%%%%% Introduction %%%%%
\section{Introduction}\label{sec:introduction}
\IEEEPARstart{F}{uture} wireless communication systems are required to support time-sensitive applications, such as robot control and automated driving, which involve sporadic data traffic due to massive short-packet communications\cite{6g_summit,MassiveAccess_WC,MassiveAccess_JSAC}. 
The conventional multiple access schemes employing grant-based approaches are unable to support such scenarios due to the non-negligible delay induced by the handshake procedure for grant acquisition\cite{GFNOMA_IoT}. 

As a promising solution to this problem, \emph{grant-free} random access schemes such as \gls{GFNOMA} in which the \gls{BS} does not exclusively assign radio resources to active users for data transmission have attracted significant attention \cite{GFNOMA_IoT,LiuMag2018}. In a \gls{GFNOMA} system, each active user can directly transmit its packet to the \gls{BS} without waiting for permission; as a result, however, the \gls{BS} must cope with \gls{AUD} to recover transmitted data accurately. 

In this context, many approaches to accurate \gls{AUD} have been investigated for synchronous \gls{GFNOMA} systems, where perfect synchronization between the \gls{BS} and users in both time and frequency is assumed\cite{Haghighatshoar2018,Liu2018-1,Senel2018mMTC,Shao2019dimension,Ke2020,Han2020,Iimori2021bilinear,Hara2022WCL,Zhu2024}. 
%%
% ここに\cite{GFNOMA_IoT}しつつ、同期が難しく、narrowband IoTなどでは周波数同期が問題となることを述べる
On the other hand, achieving ideal synchronization is impractical due to the nature of grant-free transmission, and \gls{AUD} for asynchronous scenarios remains an ongoing research area\cite{GFNOMA_IoT}. 
In practice, due to practical limitations of crystal oscillators\cite{Xu2018}, the resulting \glspl{CFO} among the users degrade the performance.

%%
% ここから、CFOが存在する状況でのAUDをレビュー(Yang Li、自分の、MLE、上田くん、Sunの)
To overcome this impediment, some approaches to \gls{AUD} in the presence of \glspl{CFO} have recently been proposed\cite{Li2019,Hara2020ICC,Liu2022ICC,Sun2022,Ueda2022PIMRC}. 
For instance, \cite{Li2019} proposed the \gls{BCD}-based method for \gls{AUD} and \gls{CE}. 
In addition, \cite{Hara2020ICC} and \cite{Liu2022ICC} handled \gls{AUD} using \gls{MLE}, which achieves higher accuracy than the \gls{BCD}-based method of \cite{Li2019}. 
Meanwhile, since the \gls{MLE}-based method uses a discrete approximation of the rotation caused by \glspl{CFO}, the improvement of the performance requires the increase in the discretization size, resulting in high computational complexity. 
As an alternative, the \gls{AMP}-based approaches for \gls{AUD} and \gls{CE} have been proposed in \cite{Sun2022} and \cite{Ueda2022PIMRC}, assuming that the frequency error is limited to a certain range that meets standard requirements. 
Although they can perform \gls{AUD} and \gls{CE} efficiently, prior information such as \emph{activity ratio} is necessary. 
This implies that a scheme to estimate prior information is required to achieve efficient \gls{AUD} in the presence of \glspl{CFO}. 

For synchronous scenarios, schemes to estimate activity ratio, namely the number of active users, have been proposed\cite{Shao2019dimension,Han2020,Zhu2024}. 
Throughout this paper, this process is referred to as \gls{AUE}. 
In \cite{Shao2019dimension}, the number of active users is estimated through rank estimation of the covariance matrix of the received pilot signal, to achieve precise \gls{AUD} using the Riemannian optimization. 
However, this approach necessitates the eigenvalue decomposition of the received pilot signal and a sufficiently long pilot sequence, resulting in high computational complexity for \gls{AUE}.
On the other hand, \cite{Han2020} and \cite{Zhu2024} have proposed the \gls{AUE} schemes using a very short common pilot sequence. 
The scheme in \cite{Han2020} uses a sequence orthogonal to the common sequence, while the scheme in \cite{Zhu2024} considers the \gls{MLE} problem of the number of active users based on the covariance matrix of the received signal. 
These \gls{AUE} schemes, utilizing a short common pilot, can estimate the number of active users with low complexity. 
\textcolor{black}{However, since these \gls{AUE} schemes use the pilot sequence, their estimation accuracy may degrade when \gls{CFO} is present, similar to \gls{AUD}.}

To this end, in this letter, we propose two \gls{AUE} schemes using a short common pilot sequence \textcolor{black}{to estimate} the number of active users in the presence of \glspl{CFO}. 
\textcolor{black}{
Although the proposed schemes use the short common pilot as in \cite{Han2020} and \cite{Zhu2024}, they do not rely on the orthogonality of the sequences or the formulation based on the \emph{pure} pilot, both of which are affected by \glspl{CFO}.
Specifically, the proposed \gls{AUE} schemes utilize the eigenvalues of the sample covariance matrix of the received signal, with one scheme designed for scenarios where the noise variance information is known, and the other for when it is unknown. 
Furthermore, the computational complexity for these schemes is comparable to that of conventional approaches.}
%%
%Specifically, these \gls{AUE} schemes that use the eigenvalues of the sample covariance matrix of the received signal, with one scheme designed for scenarios where the noise variance information is known, and the other for when it is unknown. 
%%
Numerical results demonstrate that the proposed \gls{AUE} schemes achieve superior estimation accuracy compared to the conventional schemes proposed in \cite{Han2020} and \cite{Zhu2024}. 

%% Notation
\emph{Notation:} %Throughout this paper, we use the following notation. 
The transpose and conjugate transpose are denoted by $(\cdot)^{\mathrm{T}}$ and $(\cdot)^{\mathrm{H}}$, respectively. The $l_{p}$-norm of a vector is expressed as $\|\cdot\|_{p}$. \textcolor{black}{$\Re\{\cdot\}$ and $(\cdot)^{\ast}$ denote the real part and conjugate of the argument, respectively.} In addition, $\mathbf{0}_{L}$ and $\mathbf{I}_{L}$ denote the $L$-dimensional column vectors in which all elements are zeros and the $L\times L$ identity matrix, respectively. 
The determinant and trace of a square matrix $\mathbf{A}$ are denoted by $|\mathbf{A}|$ and $\mathrm{tr}\left(\mathbf{A}\right)$. 
The expectation operator and operation that rounds $x$ to the nearest integer are represented by $\mathbb{E}[\cdot]$ and $\lfloor x \rceil$, respectively. 
\textcolor{black}{$\mathbf{a}\circ\mathbf{b}$ denotes the element-wise product of two vectors $\mathbf{a}$ and $\mathbf{b}$.}
Finally, $\mathcal{CN}(\boldsymbol{\mu},\boldsymbol{\Sigma})$ represents a multivariate complex Gaussian distribution with mean $\boldsymbol{\mu}$ and covariance $\boldsymbol{\Sigma}$. 

%%%%% System Model %%%%%
\section{System Model}\label{sec:model}
We consider an uplink \gls{GFNOMA} system comprising $N$ single antenna users and a common \gls{BS} equipped with $M$ antennas. 
At each coherence time, only $K\ll N$ users are active to transmit the packets including the same pilot used for estimating $K$ at the \gls{BS}. 
In this paper, we use $\mathbf{s}=[1\ 1]^{\mathrm{T}}$ as the same pilot. 
Furthermore, each user is assumed to be equipped with a low-cost crystal oscillator, which results in the \gls{CFO} between the users and the common \gls{BS}. 
Thus, the angular frequency between user $n$ and the \gls{BS} is modeled as $\omega_{n}\triangleq 2\pi\epsilon_{n}=2\pi\Delta f_{n}T_{\mathrm{s}}$, where $\epsilon_{n}$, $\Delta f_{n}$, and $T_{\mathrm{s}}$ are the normalized \gls{CFO}, the frequency offset in Hz, and the sampling period, respectively. 
\textcolor{black}{We assume that $\epsilon_{n}\in[-\epsilon_{\max},\epsilon_{\max}]$, where $\epsilon_{\max}$ denotes the maximum normalized \gls{CFO}\cite{Sun2022}.} 
%We model the normalized \gls{CFO} $\epsilon_{n}$ as a random variable following a uniform distribution with the interval $[-\epsilon_{\max},\epsilon_{\max}]$, where $\epsilon_{\max}$ denotes the maximum normalized \gls{CFO}\cite{Sun2022}. 

Let $\mathbf{Y}\in\mathbb{C}^{2\times M}$ denote the received pilot signal at the \gls{BS}. Then, the received signal can be written as
\begin{align}
    \mathbf{Y}&=\sum_{n\in\mathcal{A}}\sqrt{\beta_{n}\rho_{n}}(\mathbf{s}\circ\boldsymbol{\tau}(\omega_{n}))\mathbf{h}^{\mathrm{T}}_{n}+\mathbf{Z}\nonumber \\
    &=\sum_{n\in\mathcal{A}}\sqrt{\beta_{n}\rho_{n}}\boldsymbol{\tau}(\omega_{n})\mathbf{h}^{\mathrm{T}}_{n}+\mathbf{Z}, \label{eq:receive1}
\end{align}
where $\boldsymbol{\tau}(\omega_{n})=[1\ e^{j\omega_{n}}]^{\mathrm{T}}\in\mathbb{C}^{2\times 1}$ is the phase rotation vector due to $\omega_{n}$, and $\mathcal{A}\subset\{1,2,\ldots,N\}$, whose cardinality is $K$, is the set of active users. 
In addition, $\mathbf{h}_{n}\sim\mathcal{CN}(\mathbf{0}_{M},\mathbf{I}_{M}))$ and $\mathbf{Z}\in\mathbb{C}^{2\times M}$ denote the small-scale fading channel vector between user $n$ and \gls{BS}, and the noise matrix whose each entry follows an \gls{iid} complex Gaussian distribution with zero mean and variance $\sigma^{2}_{\mathrm{z}}$, respectively. $\rho_{n}$ and $\beta_{n}$ denote the transmit power of user $n$ and the large-scale fading component between user $n$ and the \gls{BS}. 
Throughout this paper, each active user is assumed to compensate for the effect of the large-scale fading component by the uplink power control, i.e., $\beta_{n}\rho_{n}=1$\cite{Han2020}. 
Thus, the received pilot signal model can be simplified as
\begin{align}
    \mathbf{Y}=\sum_{n\in\mathcal{A}}\boldsymbol{\tau}(\omega_{n})\mathbf{h}^{\mathrm{T}}_{n}+\mathbf{Z}. \label{eq:receive2}
\end{align}

Our goal is to estimate the number of active users, $K$, from the received pilot signal taking the effect of \glspl{CFO} into account. 

%%%%% Conventional AUE Schemes %%%%%
\section{Conventional AUE Schemes}
In this section, to clarify the difference between conventional and proposed \gls{AUE} schemes, we review the \gls{AUE} schemes proposed in \cite{Han2020} and \cite{Zhu2024}. 
%These conventional schemes utilize the assumption that the probability function of $\mathbf{Y}$ given $K$ and $\mathbf{s}$ is given by
\textcolor{black}{These conventional schemes operate under the assumption that the probability function of $\mathbf{Y}$ given $K$, $\mathbf{s}$, and $\omega_{n}=0, \forall n$, is given by}
\begin{align}
    p(\mathbf{Y}|K,\mathbf{s})=\frac{1}{|\pi\boldsymbol{\Sigma}|^{M}}\prod^{M}_{m=1}\exp\left(-\mathbf{y}^{\mathrm{H}}_{m}\boldsymbol{\Sigma}^{-1}\mathbf{y}^{\mathrm{H}}_{m}\right),
\end{align}
where \textcolor{black}{$\boldsymbol{\Sigma}=\mathbb{E}\left[\mathbf{y}_{m}\mathbf{y}^{\mathrm{H}}_{m}\right]=K\mathbf{s}\mathbf{s}^{\mathrm{H}}+\sigma^{2}_{\mathrm{z}}\mathbf{I}_{2}.$ Thus, this assumption does not consider the effect of \glspl{CFO}.}
% \begin{align}
%     \boldsymbol{\Sigma}=\mathbb{E}\left[\mathbf{y}_{m}\mathbf{y}^{\mathrm{H}}_{m}\right]=K\mathbf{s}\mathbf{s}^{\mathrm{H}}+\sigma^{2}_{\mathrm{z}}\mathbf{I}_{2}. \label{eq:cov1}
% \end{align}
%Note that this assumption does not consider the effect of \glspl{CFO}. 

The \gls{AUE} scheme in \cite{Han2020} utilizes the sequence $\mathbf{s}_{\perp}\in\mathbb{C}^{2\times 1}$ that is orthogonal to $\mathbf{s}$ and satisfies $\|\mathbf{s}_{\perp}\|_{2}=\|\mathbf{s}\|_{2}$. 
Let $\mathbf{R}\triangleq\mathbf{Y}\mathbf{Y}^{\mathrm{H}}/M\in\mathbb{C}^{2\times 2}$ be the sample covariance matrix of the received pilot signal in \eqref{eq:receive2}. 
Then, the criterion to estimate $K$ in \cite{Han2020} can be expressed as
\begin{align}
    \hat{K}_{\mathrm{orth}}=\Bigg\lfloor\frac{\mathbf{s}^{\mathrm{H}}\mathbf{R}\mathbf{s}}{\|\mathbf{s}\|^{4}_{2}}-\frac{\mathbf{s}^{\mathrm{H}}_{\perp}\mathbf{R}\mathbf{s}_{\perp}}{\|\mathbf{s}\|^{2}_{2}\|\mathbf{s}_{\perp}\|^{2}_{2}}\Bigg\rceil. \label{eq:conv1}
\end{align}
\textcolor{black}{When $\|\mathbf{s}\|_{2}=\|\mathbf{s}_{\perp}\|_{2}=1$, \eqref{eq:conv1} is equivalent to \cite[eq. (5)]{Han2020}. 
It is worth noting that this scheme does not require the knowledge of noise variance $\sigma^{2}_{\mathrm{z}}$.}
% When we set $\|\mathbf{s}\|_{2}=\|\mathbf{s}_{\perp}\|_{2}=1$, \eqref{eq:conv1} is equivalent to \cite[eq. (5)]{Han2020}. 
% It is worth noting that this scheme does not require the knowledge of noise variance and that $\hat{K}_{\mathrm{orth}}$ converges to $K$ when $M\to\infty$ and $\omega_{n}=0$. 

On the other hand, the \gls{AUE} scheme in \cite{Zhu2024} considers the \gls{MLE} problem of $K$. This problem can be formulated as
\begin{align}
    \underset{K\in[0,N]}{\mathrm{minimize}}&\quad\log|\boldsymbol{\Sigma}|+\mathrm{tr}\left(\boldsymbol{\Sigma}^{-1}\mathbf{R}\right) \label{eq:conv2_opt}%\\
    %\mathrm{s.t.}&\quad 0\leq K \leq N.
\end{align}
The solution of the optimization problem can be obtained by
\begin{align}
    \hat{K}_{\mathrm{MLE}}=\Bigg\lfloor\frac{\mathbf{s}^{\mathrm{H}}\mathbf{R}\mathbf{s}}{\|\mathbf{s}\|^{4}_{2}}-\frac{\sigma^{2}_{\mathrm{z}}}{\|\mathbf{s}\|^{2}_{2}}\Bigg\rceil. \label{eq:conv2}
\end{align}
\textcolor{black}{As is obvious from \eqref{eq:conv2}, the estimation of $K$ in \cite{Zhu2024} does not rely on $\mathbf{s}_{\perp}$, while it requires $\sigma^{2}_{\mathrm{z}}$.}
%As is obvious from \eqref{eq:conv2}, the estimation of $K$ in \cite{Zhu2024} does not rely on the sequence orthogonal to $\mathbf{s}$, while it requires the knowledge of noise variance. 

%%%%% Proposed AUE Schemes %%%%%
\section{Proposed AUE Schemes}\label{sec:prop}
% In this section, we propose two \gls{AUE} schemes using the sample covariance matrix of the received signal. 
\textcolor{black}{In this section, we propose two \gls{AUE} schemes, named \emph{Eig-sum} and \emph{Eig-diff}, which utilize the eigenvalues of the sample covariance matrix $\mathbf{R}$, $\lambda_{\max}$ and $\lambda_{\min}\ (<\lambda_{\max})$.}
%Specifically, the proposed schemes utilize the eigenvalues of $\mathbf{R}$. Hereafter, $\lambda_{\max}$ and $\lambda_{\min}\ (<\lambda_{\max})$ denote the eigenvalues. 
%\textcolor{black}{For convenience, we hereafter name two proposed \gls{AUE} schemes \emph{Eig-sum} and \emph{Eig-diff}, respectively.} 

\subsection{Preliminaries}
% covariance matrix with fixed $\omega_{n}$
The proposed schemes consider the covariance matrix of $\mathbf{y}_{m}$ given $\omega_{n}$ for $n\in\mathcal{A}$, i.e., 
\begin{align}
    \tilde{\boldsymbol{\Sigma}}=\mathbb{E}\left[\mathbf{y}_{m}\mathbf{y}^{\mathrm{H}}_{m}\right]&=\sum_{n\in\mathcal{A}}\boldsymbol{\tau}(\omega_{n})\boldsymbol{\tau}(\omega_{n})^{\mathrm{H}}+\sigma^{2}_{\mathrm{z}}\mathbf{I}_{2} \nonumber \\
    &=\begin{bmatrix}
    K+\sigma^{2}_{\mathrm{z}} & \sum_{n\in\mathcal{A}}e^{-j\omega_{n}} \\
    \sum_{n\in\mathcal{A}}e^{j\omega_{n}} & K+\sigma^{2}_{\mathrm{z}} \\
    \end{bmatrix} \label{eq:covariance}
\end{align}
The eigenvalues of this covariance matrix are given by
\begin{align}
    \Lambda_{\max}&=K+\sigma^{2}_{\mathrm{z}}+\gamma, \label{eq:eig_max}\\
    \Lambda_{\min}&=K+\sigma^{2}_{\mathrm{z}}-\gamma, \label{eq:eig_min}
    %\Lambda_{\max}&=K+\sigma^{2}_{\mathrm{z}}+\sqrt{\left(\sum_{n\in\mathcal{A}}e^{j\omega_{n}}\right)\left(\sum_{n\in\mathcal{A}}e^{-j\omega_{n}}\right)}, \label{eq:eig_max}\\
    %\Lambda_{\min}&=K+\sigma^{2}_{\mathrm{z}}-\sqrt{\left(\sum_{n\in\mathcal{A}}e^{j\omega_{n}}\right)\left(\sum_{n\in\mathcal{A}}e^{-j\omega_{n}}\right)}. \label{eq:eig_min}
\end{align}
where
\begin{align}
    \gamma=\sqrt{\left(\sum_{n\in\mathcal{A}}e^{j\omega_{n}}\right)\left(\sum_{n\in\mathcal{A}}e^{-j\omega_{n}}\right)}. \label{eq:gamma}
\end{align}
%The proposed \gls{AUE} schemes are designed taking this relation into account. 

Assuming $K$ is sufficiently large, $\gamma$ can be approximated as
\begin{align}
    \gamma&=K\sqrt{\left(\frac{1}{K}\sum_{n\in\mathcal{A}}e^{j\omega_{n}}\right)\left(\frac{1}{K}\sum_{n\in\mathcal{A}}e^{-j\omega_{n}}\right)} \nonumber \\
    &\approx K\sqrt{\mathbb{E}\left[e^{j\omega}\right]\mathbb{E}\left[e^{-j\omega}\right]}, \label{eq:gamma_approx}
\end{align}
\textcolor{black}{where $\omega$ represents a random variable that takes values within the interval $[-2\pi\epsilon_{\max},2\pi\epsilon_{\max}]$. $\mathbb{E}\left[e^{j\omega}\right]$ in \eqref{eq:gamma_approx} is the characteristic function of $\omega$.}
%where $\omega$ represents the random variable following a uniform distribution in the interval $[-2\pi\epsilon_{\max},2\pi\epsilon_{\max}]$. 
%%
%It is worth noting that $\mathbb{E}\left[e^{j\omega}\right]$ in \eqref{eq:gamma_approx} is the characteristic function of $\omega$, which is given by
% $\mathbb{E}\left[e^{j\omega}\right]$ in \eqref{eq:gamma_approx} is the characteristic function of $\omega$, which is given by
% \begin{align}
%     \mathbb{E}\left[e^{j\omega}\right]&=\frac{1}{4\pi\epsilon_{\max}}\int^{2\pi\epsilon_{\max}}_{-2\pi\epsilon_{\max}}e^{j\omega}d\omega \nonumber \\
%     &=\frac{\sin(2\pi\epsilon_{\max})}{2\pi\epsilon_{\max}}\nonumber \\
%     &=\mathrm{sinc}(2\pi\epsilon_{\max}). 
% \end{align} 

\subsection{\textcolor{black}{Eig-sum: Proposed AUE With Knowledge of $\sigma^{2}_{\mathrm{z}}$}}
% \subsection{Proposed AUE Scheme With Knowledge of $\sigma^{2}_{\mathrm{z}}$}
As is obvious from \eqref{eq:eig_max} and \eqref{eq:eig_min}, $K$ can ideally be obtained based on the sum of the eigenvalues of $\tilde{\boldsymbol{\Sigma}}$, i.e., $K=(\Lambda_{\max}+\Lambda_{\min})/2-\sigma^{2}_{\mathrm{z}}$, if the information of $\sigma^{2}_{\mathrm{z}}$ is available at the \gls{BS}. 
Following this fact, \textcolor{black}{Eig-sum} obtains the estimate of $K$ by
\begin{align}
    \textcolor{black}{\hat{K}_{\mathrm{sum}}=\Bigg\lfloor\frac{\lambda_{\max}+\lambda_{\min}}{2}-\sigma^{2}_{\mathrm{z}}\Bigg\rceil.} \label{eq:prop1}
\end{align}
% \begin{align}
%     \hat{K}_{\mathrm{prop1}}=\frac{\lambda_{\max}+\lambda_{\min}}{2}-\sigma^{2}_{\mathrm{z}}. \label{eq:prop1}
% \end{align}
Since this scheme is based on the calculation to offset the term including \glspl{CFO}, $\gamma$, it is expected to be robust against \glspl{CFO}. 
\par
{\color{black}
As in \cite{Han2020}, we consider the \gls{NRMSE} of this scheme, which is defined as
\begin{align}
    \textcolor{black}{\mathrm{NRMSE}=\frac{1}{K}\sqrt{\mathbb{E}\left[\left(\hat{K}-K\right)^{2}\right]},} \label{eq:NRMSE}
\end{align}
where $\hat{K}$ is the estimate of the number of active users $K$. 
Given $\alpha=\mathbb{E}[e^{j\omega}]$, the \gls{NRMSE} of Eig-sum can be obtained by
\begin{align}
    \mathrm{NRMSE}=\frac{1}{K}\sqrt{\frac{K+K(K-1)\alpha^{2}+\left(K+\sigma^{2}_{\mathrm{z}}\right)^{2}}{2M}}. \label{eq:NRMSE_prop}
\end{align}
For the derivation of \eqref{eq:NRMSE_prop}, please refer to Appendix.
}

\subsection{\textcolor{black}{Eig-diff: Proposed AUE Without Knowledge of $\sigma^{2}_{\mathrm{z}}$}}
% \subsection{Proposed AUE Scheme Without Knowledge of $\sigma^{2}_{\mathrm{z}}$}
Since $\mathbb{E}[e^{j\omega}]$ is an even function, $\mathbb{E}[e^{-j\omega}]=\mathbb{E}[e^{j\omega}]$ {\color{black} is satisfied.\footnote{\textcolor{black}{In the literature, \gls{CFO} is often modeled as a random variable that follows a symmetric distribution, such as a uniform or Gaussian distribution. Hence, the characteristic function of $\omega$ is real-valued and even.}}}
Then, \eqref{eq:gamma_approx} can be written as
\begin{align}
    \textcolor{black}{\gamma\approx K|\mathbb{E}[e^{j\omega}]|.}
\end{align}
% \begin{align}
%     \gamma\approx K|{\mathrm{sinc}}(2\pi\epsilon_{\max})|.
% \end{align}
Furthermore, since $\Lambda_{\max}-\Lambda_{\min}=2\gamma$, we can obtain the estimate of the number of active users, as follows:
\begin{align}
    \textcolor{black}{\hat{K}_{\mathrm{diff}}=\Bigg\lfloor\frac{\lambda_{\max}-\lambda_{\min}}{2|\mathbb{E}[e^{j\omega}]|}\Bigg\rceil.} \label{eq:prop2}
\end{align}
% \begin{align}
%     \hat{K}_{\mathrm{prop2}}=\frac{\lambda_{\max}-\lambda_{\min}}{2|{\mathrm{sinc}}(2\pi\epsilon_{\max})|}. \label{eq:prop2}
% \end{align}
This scheme relies solely on the knowledge of $\epsilon_{\max}$, which is usually available as the frequency errors of uplink users are confined within a certain range specified by the standard. 
\par
\textcolor{black}{Note that, since the analysis of the \gls{NRMSE} of Eig-diff is very complicated, this is left for future work.}

{\color{black}
\subsection{Complexity Comparison}
In this subsection, we discuss the computational complexity required for the proposed and conventional \gls{AUE} schemes. 
Here, we express the entries of $\mathbf{R}$, as follows:
\begin{align}
\mathbf{R}\triangleq\begin{bmatrix}
    R_{1} & \tilde{R} \\
    \tilde{R}^{\ast} & R_{2} \\
\end{bmatrix}=\begin{bmatrix}
    \frac{\|\mathbf{y}_{1}\|^{2}_{2}}{M} & \frac{\mathbf{y}_{1}\mathbf{y}^{\mathrm{H}}_{2}}{M} \\
    \left(\frac{\mathbf{y}_{1}\mathbf{y}^{\mathrm{H}}_{2}}{M}\right)^{\ast} & \frac{\|\mathbf{y}_{2}\|^{2}_{2}}{M} \\
\end{bmatrix}, \label{eq:R_define}
\end{align}
where $\mathbf{y}_{1}\in\mathbb{C}^{1\times M}$ and $\mathbf{y}_{2}\in\mathbb{C}^{1\times M}$ represent the first and second rows of the received signal $\mathbf{Y}$, respectively. 
Then, $\mathbf{y}_{1}$ and $\mathbf{y}_{2}$ are given by
\begin{align}
    \mathbf{y}_{1}=\sum_{n\in\mathcal{A}}\mathbf{h}^{\mathrm{T}}_{n}+\mathbf{z}_{1},\quad\mathbf{y}_{2}=\sum_{n\in\mathcal{A}}e^{j\omega_{n}}\mathbf{h}^{\mathrm{T}}_{n}+\mathbf{z}_{2},
\end{align}
where $\mathbf{z}_{1}\in\mathbb{C}^{1\times M}$ and $\mathbf{z}_{2}\in\mathbb{C}^{1\times M}$ are the first and second rows of the noise matrix $\mathbf{Z}$. 

Given $\mathbf{s}=[1\ 1]^{\mathrm{T}}$ and $\mathbf{s}_{\perp}=[1\ -1]^{\mathrm{T}}$, the computational complexity of \gls{AUE} schemes can be summarized in TABLE~\ref{tb:complexity}. 
This table also includes the equations for \gls{AUE} schemes, which are expressed with $R_{1}$, $R_{2}$, and $\tilde{R}$; however, their derivations are omitted owing to space constraints. 
This analysis is based on the fact that $R_{1}$, $R_{2}$, and $\tilde{R}$ can be computed using the inner product of two $M$-dimensional vectors, while ignoring constant terms that can be predetermined, such as those involving $\sigma^{2}_{\mathrm{z}}$ or $\mathbb{E}[e^{j\omega}]$. 
From TABLE~\ref{tb:complexity}, Orthogonal\cite{Han2020} has the lowest complexity, whereas Eig-sum exhibits lower complexity than MLE\cite{Zhu2024}. 
Moreover, the complexity of Eig-diff is slightly higher than, but comparable to, that of MLE\cite{Zhu2024}. 
Therefore, the proposed \gls{AUE} schemes exhibit computational complexity comparable to that of existing schemes. 
\begin{table}[t]
{\color{black}
  \begin{center}
    \caption{Computational Complexity of \gls{AUE} Schemes.}
    % \vspace{-1.0ex}
    \begin{tabular}{|c|c|}
      \hline
      \gls{AUE} scheme & Number of multiplications \\ \hline
      %Eig-sum:  & $2M+3$ \\ \hline
      Eig-sum: $\displaystyle \left\lfloor\frac{R_{1}+R_{2}}{2}-\sigma^{2}_{\mathrm{z}}\right\rceil$ & $2M+3$ \\ \hline
      Eig-diff: $\displaystyle \left\lfloor\frac{\sqrt{(R_{1}-R_{2})^{2}+4\tilde{R}\tilde{R}^{\ast}}}{2\mathbb{E}[e^{j\omega}]}\right\rceil$ & $3M+7$ \\ \hline
      Orthogonal\cite{Han2020}: $\displaystyle \left\lfloor\Re\{\tilde{R}\}\right\rceil$ & $M+1$ \\ \hline
      MLE\cite{Zhu2024}: $\displaystyle \left\lfloor\frac{R_{1}+R_{2}+2\Re\{\tilde{R}\}}{4}-\frac{\sigma^{2}_{\mathrm{z}}}{2}\right\rceil$ & $3M+5$ \\ \hline
    \end{tabular}
    \label{tb:complexity}
  \end{center}
  }
  \vspace{-4.0ex}
\end{table}
}

%%%%% Numerical Results %%%%%
\section{Numerical Results}\label{sec:results}
\begin{table}[t]
  \begin{center}
    \caption{Simulation Parameters.}
    % \vspace{-1.0ex}
    \begin{tabular}{|c|c|}
      \hline
      \textcolor{black}{Number} of potential users $N$ & 100 \\ \hline
      \textcolor{black}{Number} of active users $K$ & 25 \\ \hline
      \textcolor{black}{Number} of antennas at the \gls{BS} $M$ & 32 \\ \hline
      %Sequence length $L$ & 50 \\ \hline
      Maximum normalized CFO $\epsilon_{\max}$ & 0.15 \\ \hline
      \multirow{2}{*}{\textcolor{black}{Distribution of $\omega_{n}$}} & \textcolor{black}{Uniform distribution:} \\
      & \textcolor{black}{$\omega_{n}\in[-2\pi\epsilon_{\max},2\pi\epsilon_{\max}]$} \\ \hline
      SNR $1/\sigma^{2}_{\mathrm{z}}$ & 10 [dB] \\ \hline
    \end{tabular}
    \label{tb:setup}
  \end{center}
  \vspace{-4.0ex}
\end{table}
In this section, we investigate the performance of the proposed schemes through computer simulations. 
The simulation setup basically follows TABLE~\ref{tb:setup} unless otherwise specified. 
%To assess the accuracy of \gls{AUE}, we evaluate the normalized estimation error, which is defined as 
To assess the accuracy of \gls{AUE}, we evaluate the \textcolor{black}{\gls{NRMSE} performance, which is defined by \eqref{eq:NRMSE}.}
% \begin{align}
%     E_{K}&\triangleq\mathbb{E}\left[\frac{|\hat{K}-K|}{K}\right],
% \end{align}
% where $\hat{K}$ is the estimate of the number of active users $K$. 
In the simulations, we compare the following methods:
\begin{itemize}
    % \item \emph{Proposed1}: the proposed scheme given by \eqref{eq:prop1}.
    % \item \emph{Proposed2}: the proposed scheme given by \eqref{eq:prop2}.
    \item \textcolor{black}{\emph{Eig-sum}}: the proposed scheme given by \eqref{eq:prop1}.
    \item \textcolor{black}{\emph{Eig-diff}}: the proposed scheme given by \eqref{eq:prop2}.
    \item \emph{Orthogonal}: the \gls{AUE} scheme in \cite{Han2020} given by \eqref{eq:conv1}.
    \item \emph{MLE}: the \gls{AUE} scheme in \cite{Zhu2024} given by \eqref{eq:conv2}.
    % \item \emph{Orthogonal}: the \gls{AUE} scheme in \cite{Han2020} using the sequence $\mathbf{s}_{\perp}$ given by \eqref{eq:conv1}.
    % \item \emph{MLE}: the \gls{AUE} scheme in \cite{Zhu2024} based on the \gls{MLE} given by \eqref{eq:conv2}.
\end{itemize}
The \gls{SNR} is defined as $1/\sigma^{2}_{\mathrm{z}}$. 
\textcolor{black}{In addition, all simulation results include the analyzed \gls{NRMSE} of Eig-sum in \eqref{eq:NRMSE_prop} as ``Eig-sum (theory).''}

\begin{figure}[t]
    \begin{center}
    \subfigure[Uniform CFO]{
    \includegraphics[width=0.84\hsize]{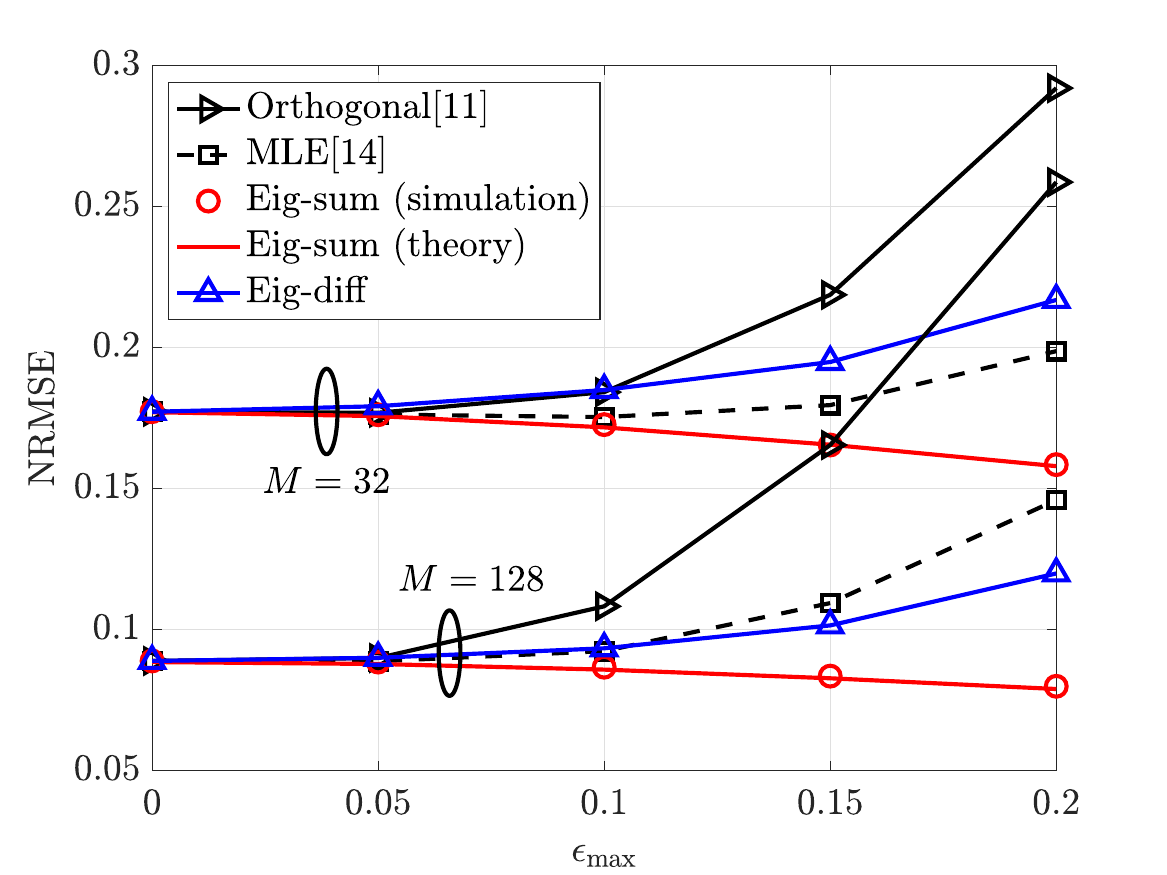}
    \label{fig:fig1_1}
    }
    \subfigure[Gaussian CFO]{
    \includegraphics[width=0.84\hsize]{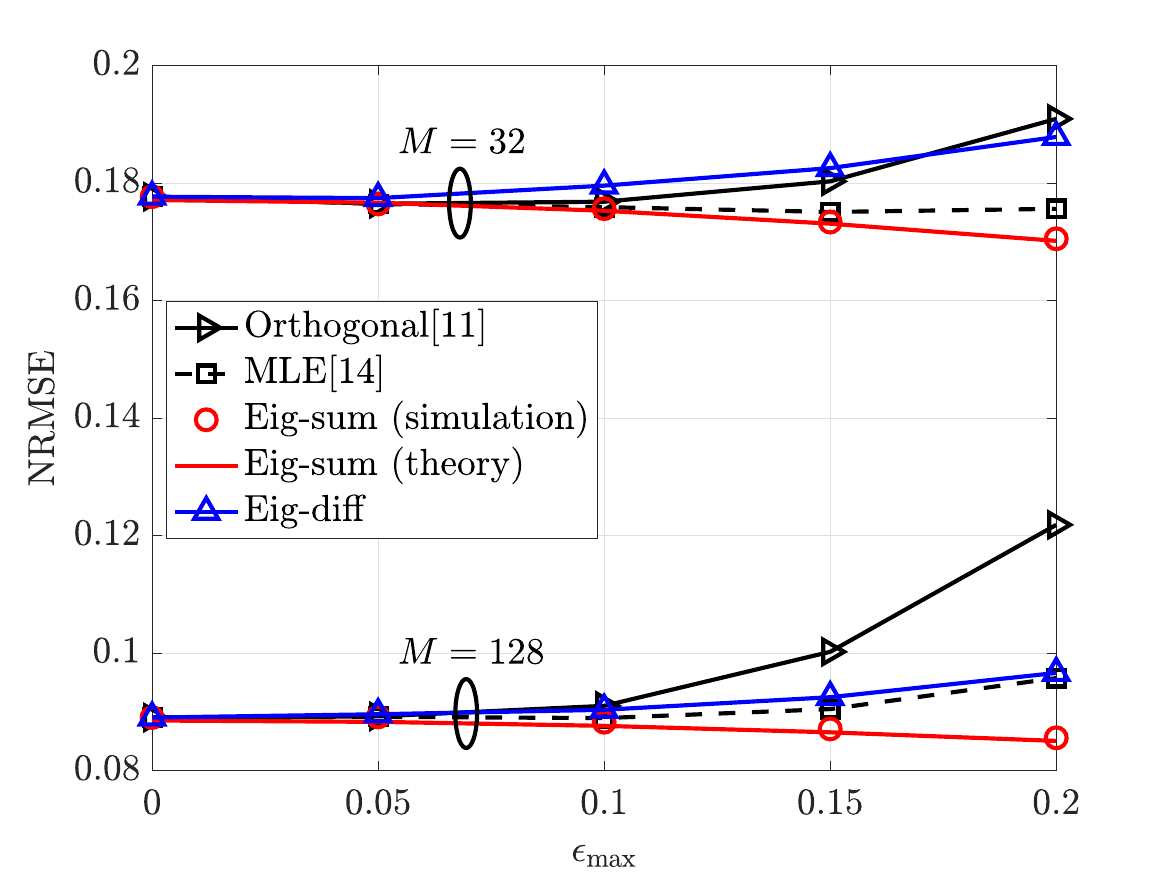}
    \label{fig:fig1_2}
    }
    \caption{\textcolor{black}{\gls{NRMSE} performance versus $\epsilon_{\max}$.}}
    \label{fig:fig1}
    \end{center}
\end{figure}

% Fig.~\ref{fig:fig1} shows the normalized estimation error performance as a function of $\epsilon_{\max}$ for $M=32$ and $128$. 
% As seen in the figure, Proposed1 outperforms the other schemes while Orthogonal performs worse than all the others, especially when $\epsilon_{\max}$ is large. 
Fig.~\ref{fig:fig1} shows the \textcolor{black}{\gls{NRMSE}} performance as a function of $\epsilon_{\max}$ for $M=32$ and \textcolor{black}{$M=128$}. 
{\color{black}
Figs.~\ref{fig:fig1_1} and \ref{fig:fig1_2} show the performance for uniform \gls{CFO} and Gaussian \gls{CFO}\footnote{\textcolor{black}{In this case, $\omega_{n}$ follows the Gaussian distribution with mean zero and standard deviation $2\pi\epsilon_{\max}/3$, in which about 99.7\% of the values lie within $[-2\pi\epsilon_{\max},2\pi\epsilon_{\max}]$.}} cases, respectively. 
As seen in Figs.~\ref{fig:fig1_1} and \ref{fig:fig1_2}, Eig-sum} outperforms the other schemes, while Orthogonal performs worse than all the others, especially when $\epsilon_{\max}$ is large. 
Interestingly, the performance of \textcolor{black}{Eig-sum} slightly improves as $\epsilon_{\max}$ increases. 
This improvement is presumed to be due to the decrease in $\gamma$ caused by the increase in $\epsilon_{\max}$. 
\textcolor{black}{Furthermore, the analyzed \gls{NRMSE} aligns with the simulated \gls{NRMSE}. We can also see that the performance of Eig-diff is superior to or comparable to that of \gls{MLE} when $M=128$, despite not using $\sigma^{2}_{\mathrm{z}}$. This is because Eig-diff considers the effect of \glspl{CFO} in the covariance matrix of the received signal, making it more robust.}
%We can also see that \textcolor{black}{Eig-diff} becomes more robust against \glspl{CFO} than \gls{MLE} when $M=128$, due to the consideration of the effect of \glspl{CFO} in the covariance matrix of the received signal. 

\begin{figure*}[t]
\centering
\begin{minipage}[b]{0.32\hsize}
    \centering
    \includegraphics[width=\hsize]{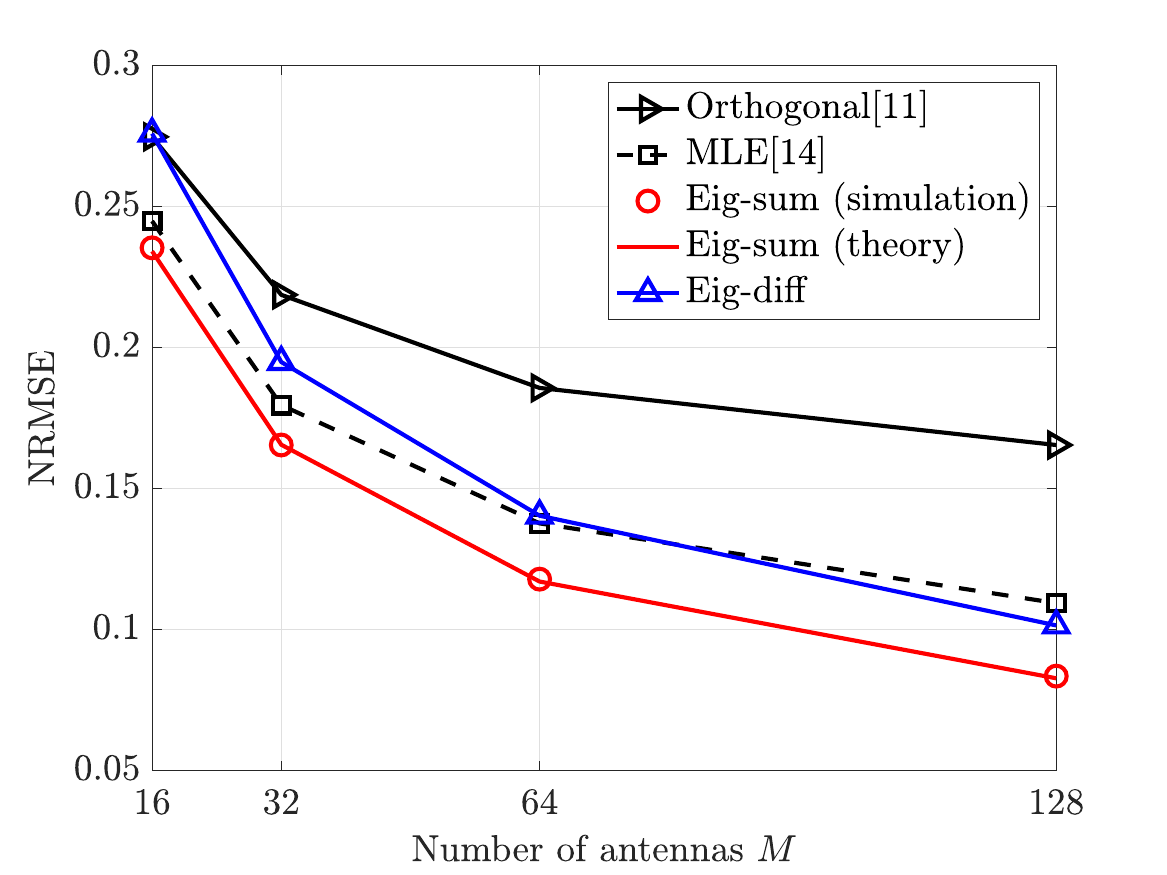}
    \caption{\textcolor{black}{\gls{NRMSE} performance versus $M$.}}
    \label{fig:fig2}
\end{minipage}
\begin{minipage}[b]{0.32\hsize}
    \centering
    \includegraphics[width=\hsize]{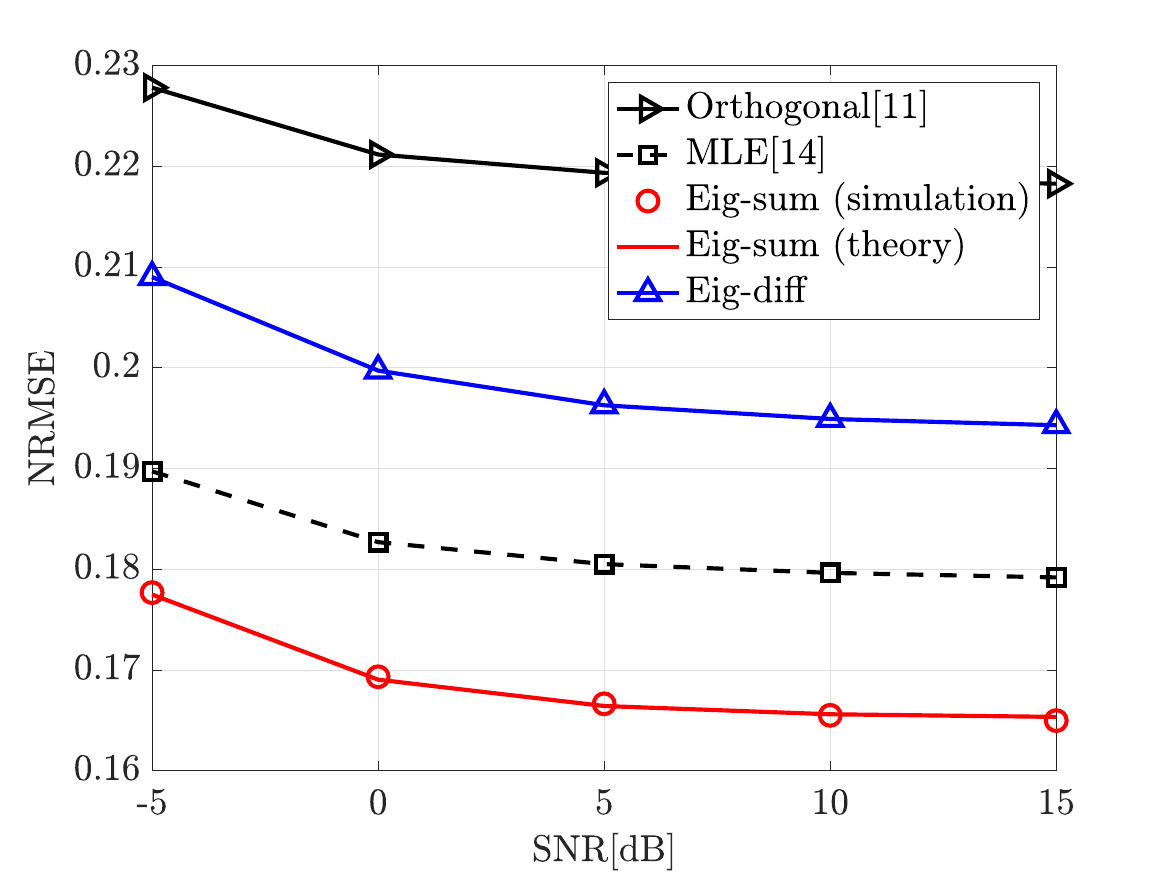}
    \caption{\textcolor{black}{\gls{NRMSE} performance versus \gls{SNR}.}}
    \label{fig:fig3}
\end{minipage}
\begin{minipage}[b]{0.32\hsize}
    \centering
    \includegraphics[width=\hsize]{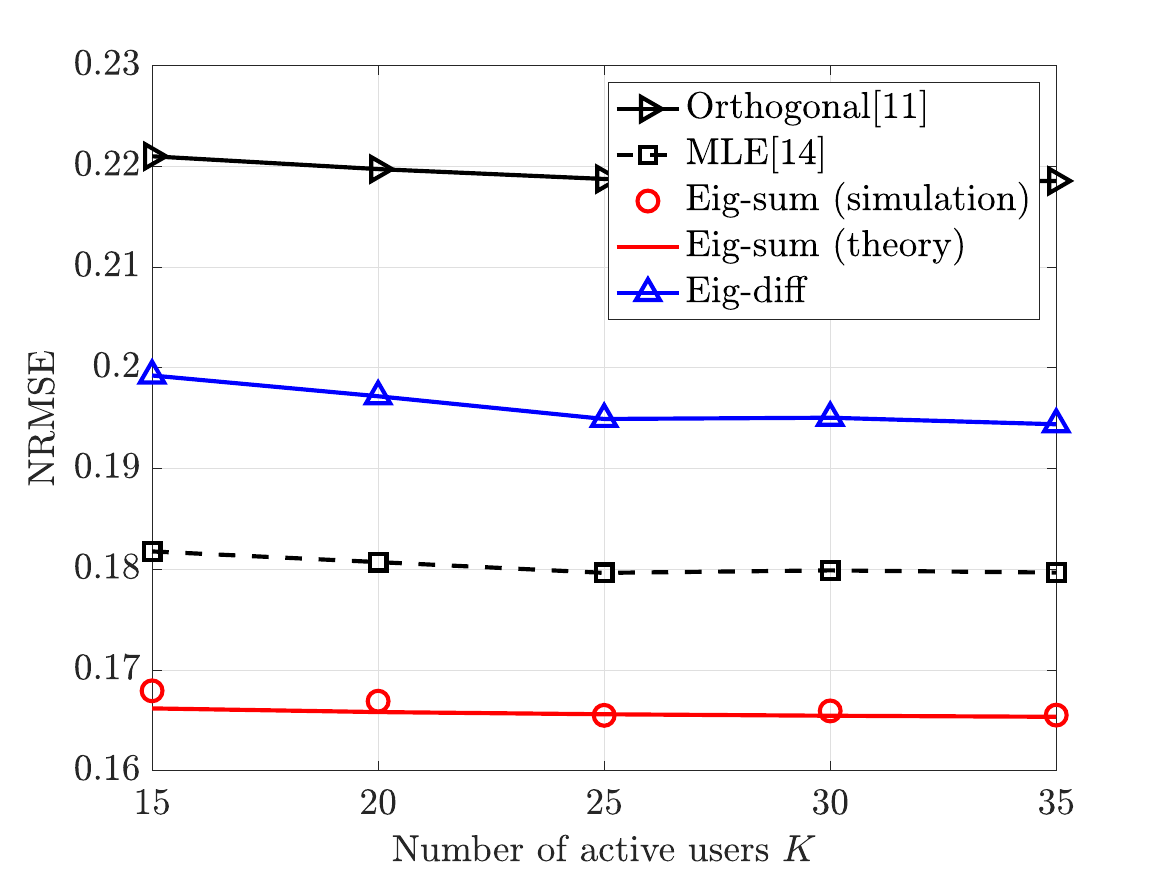}
    \caption{\textcolor{black}{\gls{NRMSE} performance versus $K$.}}
    \label{fig:fig4}
\end{minipage}
\end{figure*}

In Fig.~\ref{fig:fig2}, we show the \textcolor{black}{\gls{NRMSE}} performance when the number of antennas $M$ varies from 16 to 128. 
Similar to Fig.~\ref{fig:fig1}, this figure demonstrates that \textcolor{black}{Eig-diff performs comparably to} MLE when $M$ is sufficiently large. 
Moreover, it turns out that \textcolor{black}{Eig-sum} consistently \textcolor{black}{outperforms} the other schemes. 
% Moreover, it turns out that \textcolor{black}{Eig-sum} consistently achieves better performance than the other schemes. 
% -- subfigure ver. --

Next, we examine the effect of the noise variance $\sigma^{2}_{\mathrm{z}}$ in the \gls{AUE}. 
Fig.~\ref{fig:fig3} shows the \textcolor{black}{\gls{NRMSE}} performance versus \gls{SNR}. 
As seen in the figure, the relationship among the performances of all \gls{AUE} schemes does not vary across all \gls{SNR}. 
\textcolor{black}{Given that the \gls{NRMSE} of Eig-sum remains nearly constant with changes in \gls{SNR}, especially in the high \gls{SNR} region, it is reasonable to presume that the \gls{NRMSE} of other methods would similarly remain largely unaffected by \gls{SNR} variations.}
% In addition, \textcolor{black}{Eig-diff} follows the same performance trend as the other schemes, even though it does not use the knowledge of $\sigma^{2}_{\mathrm{z}}$. 

Finally, Fig.~\ref{fig:fig4} shows the \textcolor{black}{\gls{NRMSE}} performance versus the number of active users $K$. 
This result indicates that the estimation error of the proposed schemes remains roughly constant under different value of $K$, similar to the conventional schemes. 
In light of the above, we can conclude that the proposed schemes are robust to both \glspl{CFO} and the number of active users. 

%%%%% Conclusion %%%%%
\section{Conclusion}\label{sec:conclusion}
In this letter, we proposed two \gls{AUE} schemes for \gls{GFNOMA} in the presence of \glspl{CFO}. 
Both proposed schemes utilize the eigenvalues of the sample covariance matrix of the received signal, with one scheme designed to function without requiring noise variance information. 
Numerical results demonstrate that the proposed schemes outperform conventional methods when the noise variance is known. Moreover, even without knowledge of the noise variance, our schemes perform comparably to these conventional approaches.
\textcolor{black}{Although this study focuses on a scenario with \glspl{CFO}, addressing timing offsets is also necessary and is therefore left for future work.}

{\color{black}
\appendix
The \gls{NRMSE} of Eig-sum can be calculated by \eqref{eq:NRMSE3}, which is based on $\hat{K}_{\mathrm{sum}}=\lfloor(R_{1}+R_{2})/2-\sigma^{2}_{\mathrm{z}}\rceil$.
\begin{figure*}[b]
\hrulefill
\begin{align}
    \textcolor{black}{\mathrm{NRMSE}=\frac{1}{K}\sqrt{\mathbb{E}\left[\left(\frac{R_{1}+R_{2}}{2}-\sigma^{2}_{\mathrm{z}}\right)^2\right]-K^2}
    =\frac{1}{K}\sqrt{\frac{1}{4}\mathbb{E}\left[R^{2}_{1}\right]+\frac{1}{4}\mathbb{E}\left[R^{2}_{2}\right]+\frac{1}{2}\mathbb{E}\left[R_{1}R_{2}\right]-\sigma^{2}_{\mathrm{z}}\left(\mathbb{E}\left[R_{1}\right]+\mathbb{E}\left[R_{2}\right]\right)+\sigma^{4}_{\mathrm{z}}-K^2}.} \label{eq:NRMSE3}
\end{align}
\end{figure*}
%This is based on the relationship: $\hat{K}_{\mathrm{sum}}=\lfloor(R_{1}+R_{2})/2-\sigma^{2}_{\mathrm{z}}\rceil$. 
Note that the derivation of \eqref{eq:NRMSE_prop} ignores the rounding operation as in \cite{Han2020}. 

We first consider $\mathbb{E}[R_{1}]$, $\mathbb{E}[R_{1}]$, $\mathbb{E}[R^{2}_{1}]$, and $\mathbb{E}[R^{2}_{2}]$.
Since $\mathbf{y}_{1}$ and $\mathbf{y}_{2}$ can be regarded as a random vector that follows $\mathcal{CN}(\mathbf{0}_{M},(K+\sigma^{2}_{\mathrm{z}})\mathbf{I}_{M})$, $R_{1}$ and $R_{2}$ follow the Erlang distribution with the shape parameter $M$ and rate parameter $M/(K+\sigma^{2}_{\mathrm{z}})$. 
The mean and variance of this distribution are $K+\sigma^{2}_{\mathrm{z}}$ and $(K+\sigma^{2}_{\mathrm{z}})^{2}/M$, respectively. 
Consequently, we can obtain the following relationships. 
\begin{align}
    \mathbb{E}[R_{1}]&=\mathbb{E}[R_{2}]=K+\sigma^{2}_{\mathrm{z}}, \label{eq:R_mean} \\ 
    \mathbb{E}[R^{2}_{1}]&=\mathbb{E}[R^{2}_{2}]=\left(1+\frac{1}{M}\right)\left(K+\sigma^{2}_{\mathrm{z}}\right)^{2}. \label{eq:R_sq_mean}
\end{align}

Next, we discuss the value of $\mathbb{E}[R_{1}R_{2}]$. 
$R_{1}$ and $R_{2}$ can be expanded as follows:
\begin{align}
    R_{1}&=\sum_{n\in\mathcal{A}}\frac{\|\mathbf{h}_{n}\|^{2}_{2}}{M}+\sum_{n\in\mathcal{A}}\sum_{n'\in\mathcal{A},n'\neq n}\frac{\mathbf{h}^{\mathrm{T}}_{n}\mathbf{h}^{\ast}_{n'}}{M} \nonumber \\
    &\quad+\sum_{n\in\mathcal{A}}\frac{\mathbf{h}^{\mathrm{T}}_{n}\mathbf{z}^{\mathrm{H}}_{1}}{M}+\sum_{n\in\mathcal{A}}\frac{\mathbf{z}_{1}\mathbf{h}^{\ast}_{n}}{M}+\frac{\|\mathbf{z}_{1}\|^{2}_{2}}{M}, \label{eq:R1}\\
    R_{2}&=\sum_{n\in\mathcal{A}}\frac{\|\mathbf{h}_{n}\|^{2}_{2}}{M}+\sum_{n\in\mathcal{A}}\sum_{n'\in\mathcal{A},n'\neq n}e^{j(\omega_{n}-\omega_{n'})}\frac{\mathbf{h}^{\mathrm{T}}_{n}\mathbf{h}^{\ast}_{n'}}{M} \nonumber \\
    &\quad+\sum_{n\in\mathcal{A}}e^{j\omega_{n}}\frac{\mathbf{h}^{\mathrm{T}}_{n}\mathbf{z}^{\mathrm{H}}_{2}}{M}+\sum_{n\in\mathcal{A}}e^{-j\omega_{n}}\frac{\mathbf{z}_{2}\mathbf{h}^{\ast}_{n}}{M}+\frac{\|\mathbf{z}_{2}\|^{2}_{2}}{M}. \label{eq:R2}
\end{align}
Given that $\mathbf{h}_{n}$, $\omega_{n}$, $\mathbf{z}_{1}$, and $\mathbf{z}_{2}$ are independent of each other, $R_{1}R_{2}$ can be expressed by \eqref{eq:R1R2} without including negligible terms, 
\begin{figure*}[b]
\vspace{-3.0ex}
\begin{align}
    \textcolor{black}{R_{1}R_{2}=\!\sum_{n\in\mathcal{A}}\frac{\|\mathbf{h}_{n}\|^{4}_{2}}{M^2}+\!\sum_{n\in\mathcal{A}}\sum_{n'\in\mathcal{A},n'\neq n}\left(\frac{\|\mathbf{h}_{n}\|^{2}_{2}\|\mathbf{h}_{n'}\|^{2}_{2}}{M^2}+e^{j(\omega_{n}-\omega_{n'})}\frac{\mathbf{h}^{\mathrm{T}}_{n'}\mathbf{h}^{\ast}_{n'}\mathbf{h}^{\mathrm{T}}_{n'}\mathbf{h}^{\ast}_{n}}{M^{2}}\right)+\frac{\|\mathbf{z}_{1}\|^{2}_{2}+\|\mathbf{z}_{2}\|^{2}_{2}}{M^{2}}\!\sum_{n\in\mathcal{A}}\|\mathbf{h}_{n}\|^{2}_{2}+\frac{\|\mathbf{z}_{1}\|^{2}_{2}\|\mathbf{z}_{2}\|^{2}_{2}}{M^{2}}.} \label{eq:R1R2}
\end{align}
\end{figure*}
where we use the following relationship:
\begin{align}
    \sum_{n\in\mathcal{A}}\sum_{n'\in\mathcal{A},n'\neq n}\frac{\mathbf{h}^{\mathrm{T}}_{n}\mathbf{h}^{\ast}_{n'}}{M}=\sum_{n\in\mathcal{A}}\sum_{n'\in\mathcal{A},n'\neq n}\frac{\mathbf{h}^{\mathrm{T}}_{n'}\mathbf{h}^{\ast}_{n}}{M}.
\end{align}
In addition, since $\mathbf{h}_{n}\sim\mathcal{CN}(\mathbf{0}_{M},\mathbf{I}_{M})$, $\mathbf{z}_{1}\sim\mathcal{CN}(\mathbf{0}_{M},\sigma^{2}_{\mathrm{z}}\mathbf{I}_{M})$, and $\mathbf{z}_{2}\sim\mathcal{CN}(\mathbf{0}_{M},\sigma^{2}_{\mathrm{z}}\mathbf{I}_{M})$, we can obtain the relationships: $\mathbb{E}\left[\|\mathbf{h}_{n}\|^{4}_{2}\right]=M+M^2$, $\mathbb{E}\left[\|\mathbf{h}_{n}\|^{2}_{2}\right]=M$, $\mathbb{E}\left[\mathbf{h}^{\ast}_{n'}\mathbf{h}^{\mathrm{T}}_{n'}\right]=\mathbf{I}_{M}$, and $\mathbb{E}\left[\|\mathbf{z}_{1}\|^{2}_{2}\right]=\mathbb{E}\left[\|\mathbf{z}_{2}\|^{2}_{2}\right]=M\sigma^{2}_{\mathrm{z}}$.
As a result, $\mathbb{E}[R_{1}R_{2}]$ can be calculated by
\begin{align}
    \mathbb{E}\left[R_{1}R_{2}\right]=\frac{K+K(K-1)\alpha^{2}}{M}+\left(K+\sigma^{2}_{\mathrm{z}}\right)^{2}. \label{eq:RR_mean}
\end{align}

In light of the above, by substituting \eqref{eq:R_mean}, \eqref{eq:R_sq_mean}, and \eqref{eq:RR_mean} into \eqref{eq:NRMSE3}, we can derive \eqref{eq:NRMSE_prop}. 
}
% \begin{align}
%     R_{1}R_{2}&=\sum_{n\in\mathcal{A}}\frac{\|\mathbf{h}_{n}\|^{4}_{2}}{M^2}+\sum_{n\in\mathcal{A}}\sum_{n'\in\mathcal{A},n'\neq n}\frac{\|\mathbf{h}_{n}\|^{2}_{2}\|\mathbf{h}_{n'}\|^{2}_{2}}{M^2} \nonumber\\
%     &\quad+\sum_{n\in\mathcal{A}}\sum_{n'\in\mathcal{A},n'\neq n}e^{j(\omega_{n}-\omega_{n'})}\frac{\mathrm{tr}\left(\mathbf{h}^{\ast}_{n'}\mathbf{h}^{\mathrm{T}}_{n'}\mathbf{h}^{\mathrm{T}}_{n}\mathbf{h}^{\ast}_{n}\right)}{M^{2}} \nonumber\\
%     &\quad+\frac{\|\mathbf{z}_{1}\|^{2}_{2}+\|\mathbf{z}_{2}\|^{2}_{2}}{M^{2}}\sum_{n\in\mathcal{A}}\|\mathbf{h}_{n}\|^{2}_{2}+\frac{\|\mathbf{z}_{1}\|^{2}_{2}\|\mathbf{z}_{2}\|^{2}_{2}}{M^{2}},
% \end{align}

%%%%% References %%%%%
\bibliography{IEEEabrv,IEEEWCL2024}
\end{document}